\documentstyle[epsfig]{aipproc}

\def\lsim{\mathrel{\mathpalette\fun <}}
\def\gsim{\mathrel{\mathpalette\fun >}}
\def\fun#1#2{\lower3.6pt\vbox{\baselineskip0pt\lineskip.9pt
\ialign{$\mathsurround=0pt#1\hfil##\hfil$\crcr#2\crcr\sim\crcr}}}

\begin{document}
\title{Magnetic lensing\\of ultra high energy cosmic rays
\footnote{Prepared for the Proceedings of the International Workshop 
on Observing Ultra High Energy Cosmic Rays from Space and Earth,
Metepec, Puebla, Mexico, August 9-12, 2000.}}

\author{Diego Harari$^*$, Silvia Mollerach$^{\dagger}$ and 
Esteban Roulet$^{\dagger}$}

\address{$^*$Departamento de F\'\i sica, FCEyN, Universidad de Buenos Aires
\\Ciudad Universitaria - Pab. 1, 1428 Buenos Aires, Argentina\\
$^{\dagger}$Departamento de F\'\i sica, 
Universidad Nacional de La Plata\\CC 67, 1900 La Plata, Argentina}


\maketitle

\begin{abstract}
We discuss several effects due to lensing of ultra high 
energy cosmic rays by the regular component of the galactic 
magnetic field. Large flux magnification around caustics can be
a significant source of clustering in the arrival directions of 
UHECRs of comparable energy.
We also discuss lensing effects in a hypothetical 
galactic magnetic wind model recently proposed
to explain the extremely high energy cosmic rays 
so far observed as originating from a single source (M87).
This model implies large flux magnifications, which reduce 
the power requirements on the source, and a significant asymmetry
in the expected flux between the north and south galactic hemispheres.
 
\end{abstract}

\section*{Lensing in the regular galactic field}

Galactic and intergalactic magnetic fields play a major role in
the physics of ultra high energy cosmic rays (CRs) \cite{Bs}. 
The magnetic field of our Galaxy can lead to sizeable deflections
of CR trajectories \cite{st97,me98}.
Here we emphasize the action of the galactic magnetic field as a giant 
lens which can sizeably amplify the CR flux arriving from any given 
source in some energy range \cite{I,II}. We illustrate these effects 
within a bisymmetric spiral model for the regular component 
of the galactic magnetic field, with a value at the location of 
the solar system of 2 $\mu$G (see \cite{I,II} for details). 
Precise predictions depend upon the detailed structure of the galactic 
magnetic field, which is not so well known. The main features should however
be rather generic to any realistic model. In the case here considered
magnetic lensing is relevant in the energy range
$~5\times 10^{18}~{\rm eV} \lsim E/Z \lsim 5\times 10^{19}~{\rm eV}$, where
$Z$ is the CR electric charge.
At higher values of $E/Z$ deflections and magnifications are not very large, 
while at smaller energies the drift and diffusive regimes dominate.

A bundle of CR trajectories from an extragalactic source can be focused
as it traverses the inhomogeneous galactic magnetic field. Its amplification 
is given by the ratio of the flux of particles reaching the Earth 
to that outside the region of influence of the galactic magnetic field. 
Fig. 1 illustrates the effect in the case of CRs that arrive from the 
direction to M87 in the Virgo cluster, specified by galactic coordinates 
$(\ell,b)=(282.5^\circ,74.4^\circ)$.
\begin{figure}[b!] 
\centerline{\epsfig{file=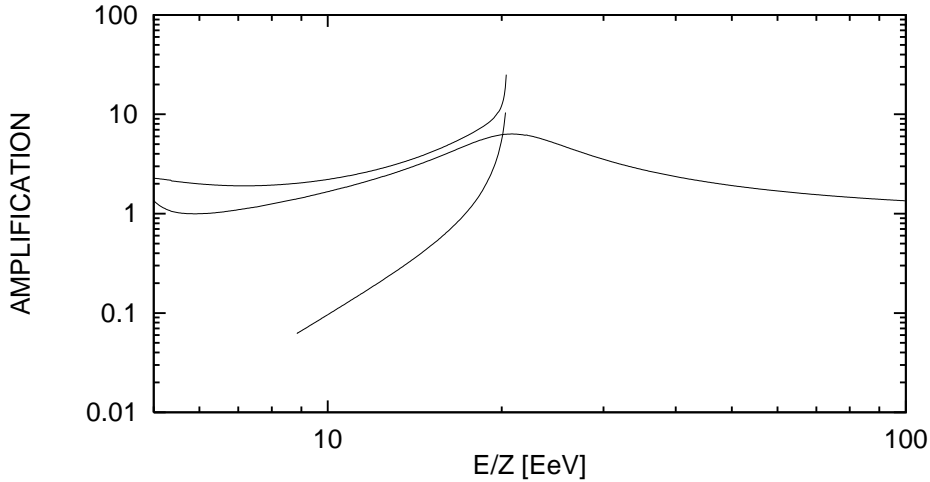,width=5in}}
\vspace{10pt}
\caption{Amplification vs. energy of the flux in the 
principal and secondary images of an extragalactic  
source in the direction of M87.}
\label{fig1}
\end{figure}
The flux from the principal image (the one visible at the highest energies) 
is amplified a few times at energies  $10~{\rm EeV} \lsim E/Z \lsim 
30~{\rm EeV}$ (1~EeV=$10^{18}~{\rm eV}$). 
At an energy $E/Z\approx 20~{\rm EeV}$ the source position
lies along a caustic of the magnetic lens mapping.
A pair of new images of the same source becomes visible from Earth at
energies below the energy of the caustic. 

The energy dependence of the
amplification of each secondary image near the caustic is fitted with 
very high accuracy as $\mu_i(E)\approx{A\over \sqrt{1-E/E_0}}\pm B
,~(i=1,2)$
(in the particular case displayed in Fig. 1, $A\approx 1.3$). 
This dependence can be intuitively understood 
as a consequence of the energy-dependence of the location of the caustics
\cite{II}. The number of events expected within a given energy bin is
obtained from the convolution of the original spectrum produced by the
source and the magnification due to the intervening magnetic field. In
view of the energy dependence of the magnification near the caustic it
is clear that in spite of its divergent behaviour, a finite number of
events results. 
Taking as an illustration an injected differential spectrum 
that scales as $E^{-2.7}$, the integrated flux from two secondary
images in the energy
interval between $0.9~E_0$ and $E_0$ turns out to be $12~A$ times 
larger than the flux of the principal 
image of the source in the same energy range in the 
absence of magnification. This is also $2.4A$ times the 
flux that would arrive from the principal image at 
all energies above $E_0$ if there were no magnetic lensing. 
Detection of an UHECR source in a narrow energy range around 
a caustic may thus be more likely than its detection at any 
higher energy. 

The enhancement of the probability to detect events from a given source  
in a narrow energy range near the caustic implies a concentration of 
events around the location at which the new pair of images forms. 
Magnetic lensing around caustics is thus a potential source of
clustering in the angular distribution of CR arrival directions, and 
should be taken into account in statistical analysis of small scale 
clustering of observed events.
In this respect it is remarkable that the observed
events in doublets and triplets are in most cases very close in
energy \cite{uchi99}. This may be an indication that at least a fraction of
the observed clustering of events may be due to magnetic
lensing around caustics. 

Caustics are not an uncommon feature, because they sweep a rather 
significant fraction of the sky as the ratio 
$E/Z$ varies between around 50~EeV and a few EeV. To illustrate this point
we display in Fig. 2 the contour plots of
the magnification of the CR flux from point sources as a function of
the CR arrival direction at Earth, for  $E/Z=30$ and  10~EeV. 
The critical curves in the amplification maps at the observer's plane
correspond to the caustics in the source plane, and are within the black
contours in Fig.~2.

\begin{figure}[b!] 
\centerline{\epsfig{file=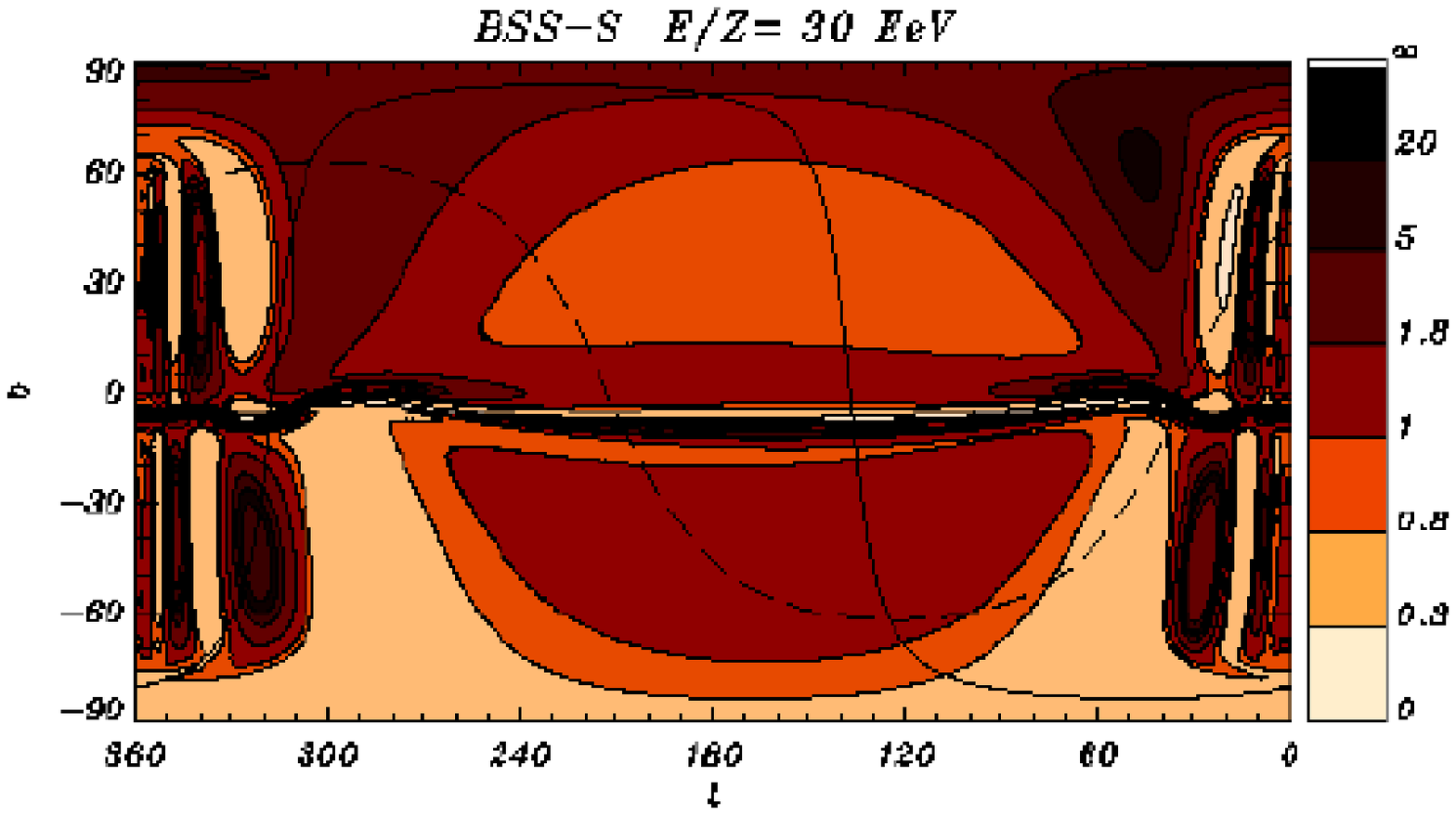,width=2.9in}
\epsfig{file=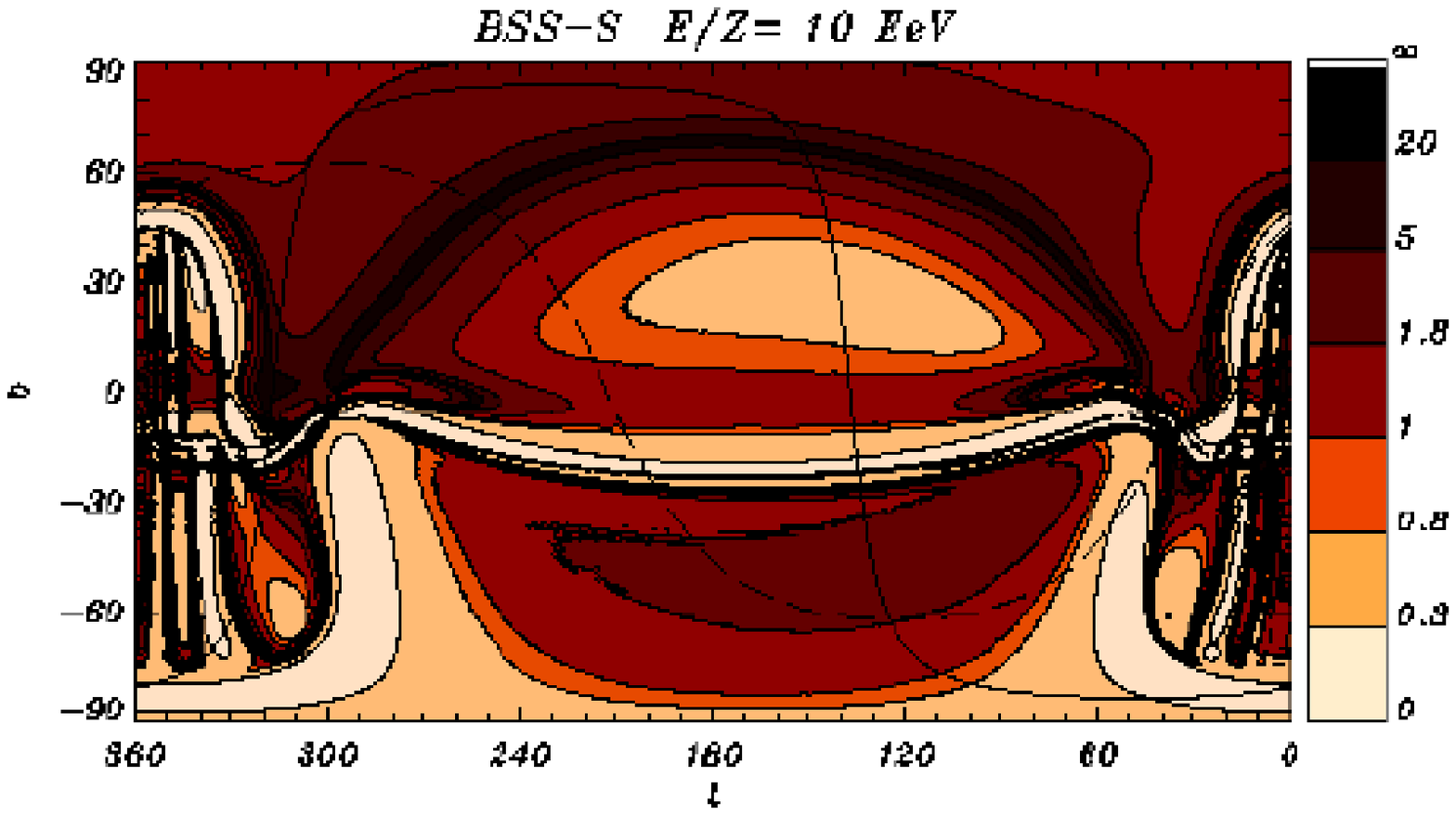,width=2.9in}}
\vspace{10pt}
\caption{Contour plots of
the magnification of the CR flux from a point source as a function of
the  arrival direction at Earth, for
$E/Z=30$ EeV (left) and 10 EeV (right).}
\label{fig2}
\end{figure}

We stress the fact that Fig. 2 plots the expected magnification for CRs
arriving to Earth from isolated point sources. If the flux incoming to
the region of influence of the galactic magnetic field were isotropic,
the observed flux would also be seen isotropic, as a consequence of
Liouville theorem. This fact was pointed out by Lema\^\i tre and 
Vallarta in their study of the deflection of low energy 
(few tenths of GeV) CRs by the Earth  dipole magnetic field,
in a paper which was quoted in a round table at this meeting
as the first publication by a mexican author in the field of CRs 
\cite{le33}. 

Other implications of lensing by the galactic magnetic field
include effects upon the observed CR composition. Nuclei with 
different charges are magnified by different amounts for a given energy,
which may lead to a sizeable effect for sources whose magnification 
has a strong energy dependence (in particular around caustics) and 
which have a mixed composition.

Another peculiar feature of lensing in the vicinity of caustics is 
that the relative arrival time of events from a single 
image of a CR source does not necessarily increase with decreasing 
energy. It is often argued that the doublets in which the 
highest energy event arrived later than the other member in the pair 
can not arise from bursting sources. This is not necessarily true 
near a caustic.

\section*{Lensing in a hypothetical galactic wind}
The angular distribution of the extreme high energy (EHE) events
($E\gsim 10^{20}$~eV) so far observed is consistent 
with isotropy. 
There are no known nearby sources close to their arrival directions 
considered to be a potential site for acceleration of cosmic rays 
to such enormous energies.
It is possible that EHECRs are protons or nuclei emitted by a few 
nearby extragalactic sources, but then intervening magnetic fields 
should significantly bend their trajectories to explain why their 
arrival directions do not point to their place of origin. 
The regular component of the galactic magnetic field does not
produce large deflections at these extremely high energies except 
over a heavy (large $Z$) component of the CR flux.

It has recently been speculated \cite{ahn99,bier00} that all the events so
far detected at energies above $10^{20}$ eV may originate from M87 in the
Virgo cluster, if the Galaxy has a rather strong and extended
magnetic wind, if two out of the thirteen events considered are He nuclei
(the rest being protons), and  if 
intergalactic magnetic fields provide the extra deflection 
(of order $20^\circ$) needed to fine-tune the incoming particles 
in the appropriate direction (close to the direction to the 
northern galactic pole \cite{bi00}) as they enter the wind.

\begin{figure}[b!] 
\centerline{\epsfig{file=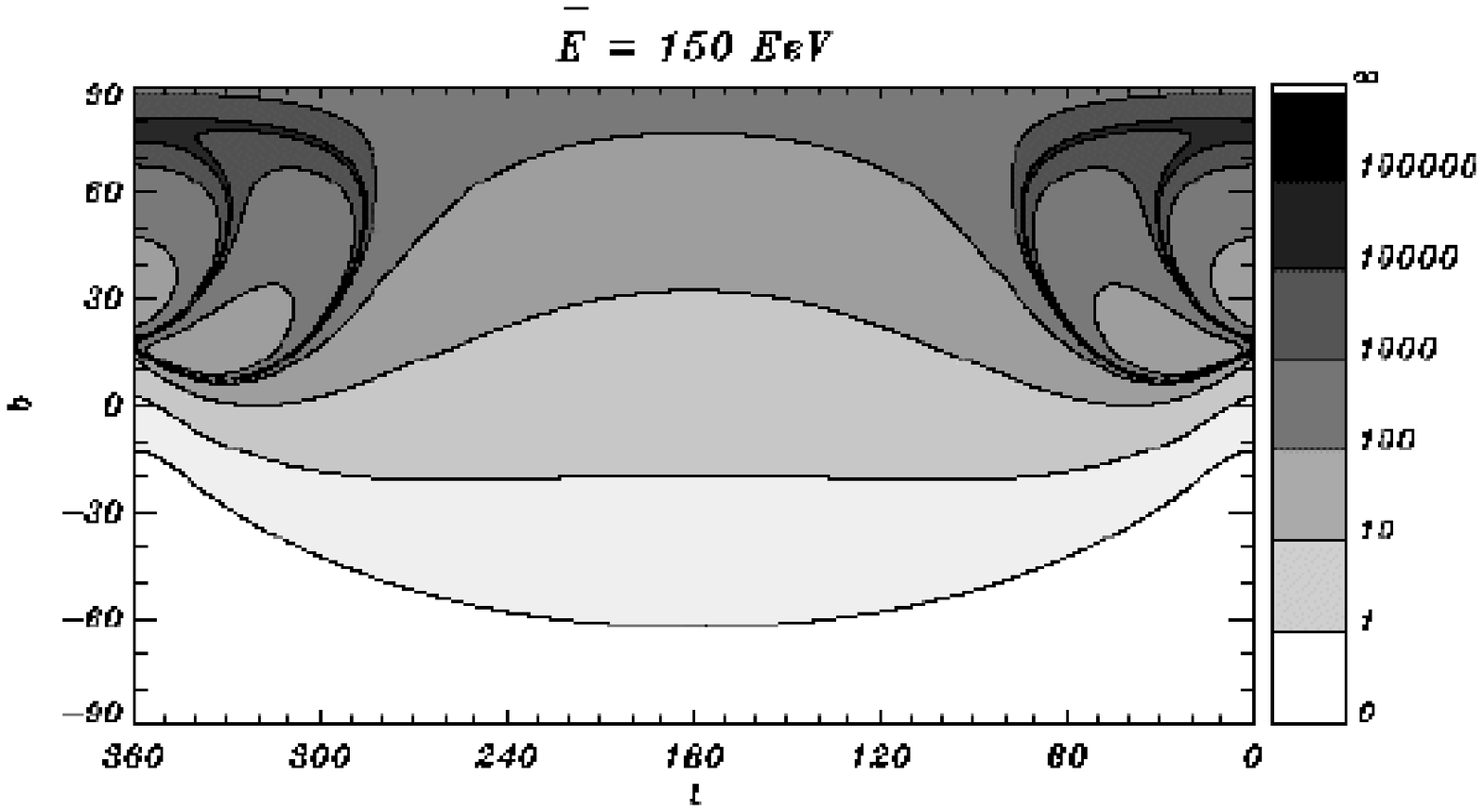,width=2.9in}
\epsfig{file=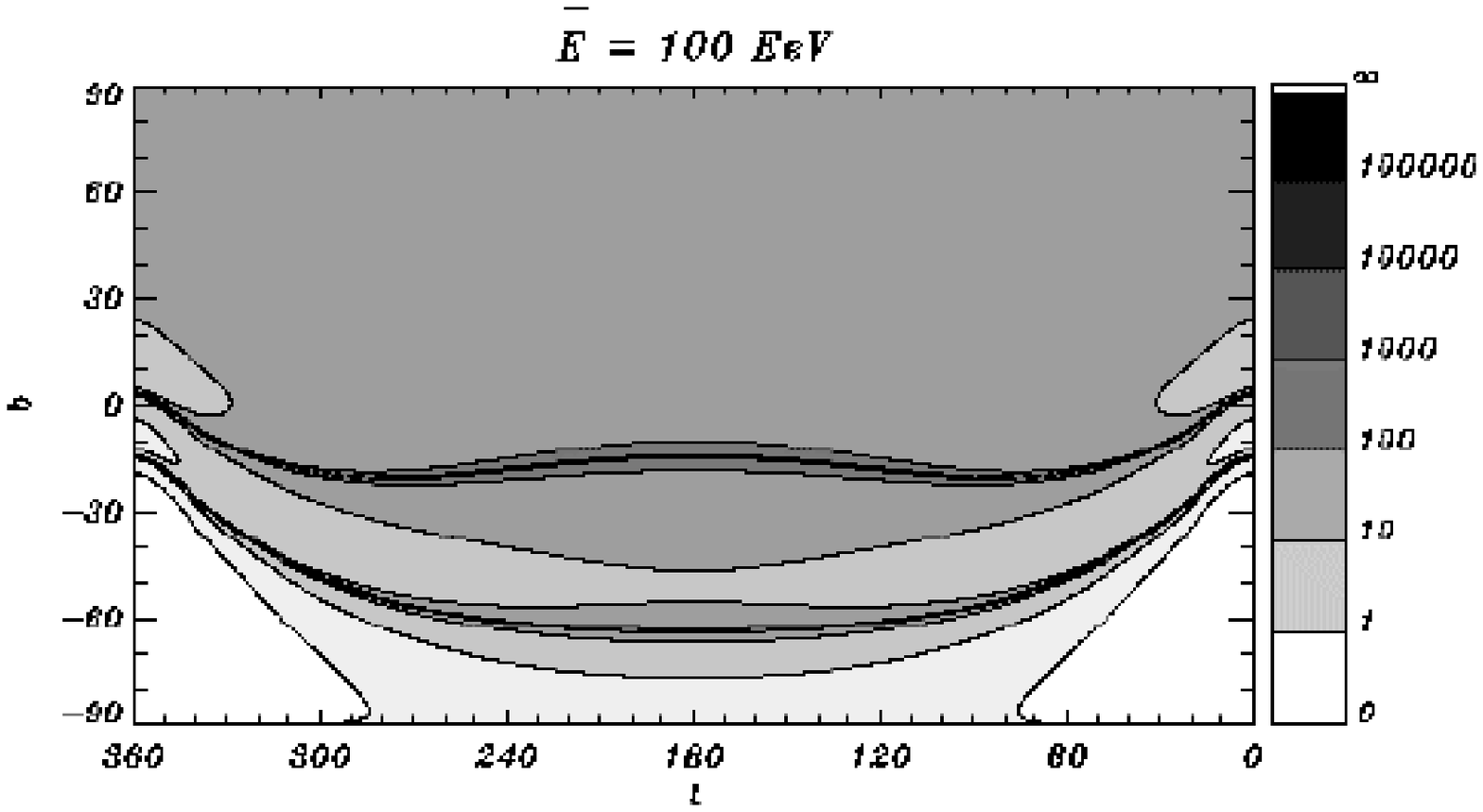,width=2.9in}}
\vspace{10pt}
\caption{Contour plots of
the magnification by 
the galactic magnetic wind of the CR flux from a point
source as a function of
the  arrival direction to the Earth  for two different 
values of  $\overline E\equiv E/(ZB_7)$.}
\label{fig3}
\end{figure}

Here we discuss some generic predictions due to flux magnification
by magnetic lensing in this wind model that may serve to test
its validity as more data becomes available\cite{bgalw}.
The idealized magnetic wind model considered is an azimuthal field 
with strength given by 
$B=B_7\ 7\mu{\rm G}{r_0\over r}\sin\theta \tanh(r/r_s)$
as a function of the radial spherical coordinate $r$ and the angle
to the north galactic pole $\theta$. The distance from the Earth
to the galactic center is 
$r_0=8.5~$kpc. $B_7$ is a normalization factor.
Lensing effects depend upon the magnetic field strength and the 
CR energy and charge only through the combination
$\overline E\equiv E/(ZB_7)$.  
The factor $\tanh(r/r_s)$ was
introduced to smooth out the field at small radii. We took
$r_s=5$~kpc. We adopted a 1.5~Mpc cutoff for the extension of the 
field. The main lensing effects are produced at distances 
larger than 10~kpc and less than a few hundred kpc, so that  
taking $r_s=10$~kpc or a cutoff at 1~Mpc leaves the 
results essentially unchanged.

Fig. 3 shows contour plots of equal magnification for given
CR arrival directions. Huge magnifications, in excess than a
factor of 100, are attained in large regions of the sky. 
Most of the sky is swept from north to south by the critical 
lines (which are also indicative of multiple image formation) 
as $\overline E$ varies between 150~EeV and just below 100~EeV.
 
CRs at different energies enter the galactic wind 
from different directions if they have suffered magnetic deflections
in their way. It is nevertheless instructive to
consider fixed incoming
directions, to illustrate some of the generic features of  
lensing by the galactic wind.
Figure~4 displays, in its left panel, the energy-dependent magnification 
of the
flux of CRs that enter the galactic magnetic wind from the direction
($\ell, b)=(270^\circ,88^\circ)$, 
for the principal (P) as
well as for the secondary images (A, B). 
The right panel of Fig. 4 shows the expected arrival directions
of 50 equally probable events from a source that injects CRs
from the same fixed direction with a differential energy spectrum
proportional to  $E^{-2.7}.$ 
\begin{figure}[b!] 
\centerline{\epsfig{file=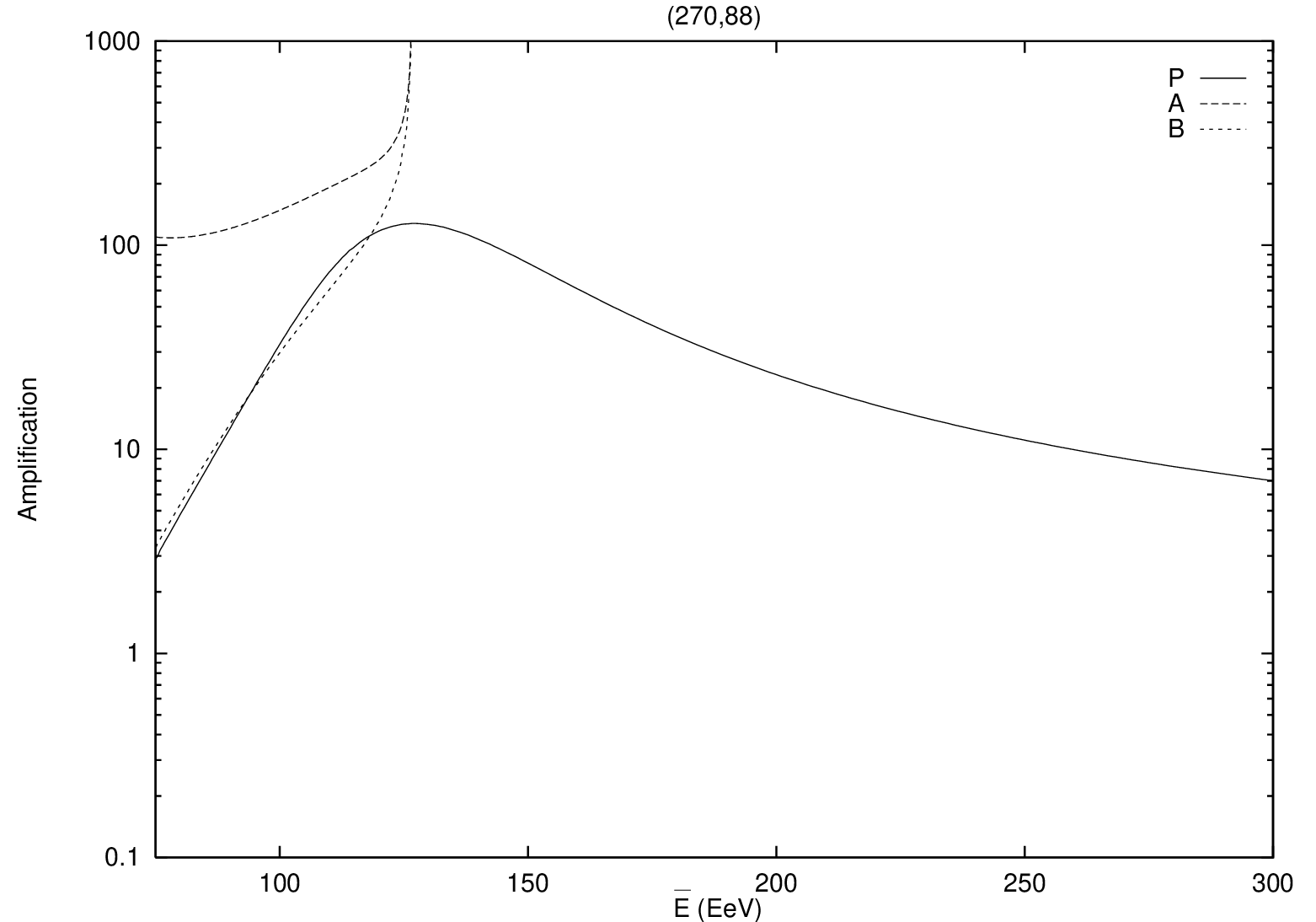,width=2.9in,angle=-180}
\epsfig{file=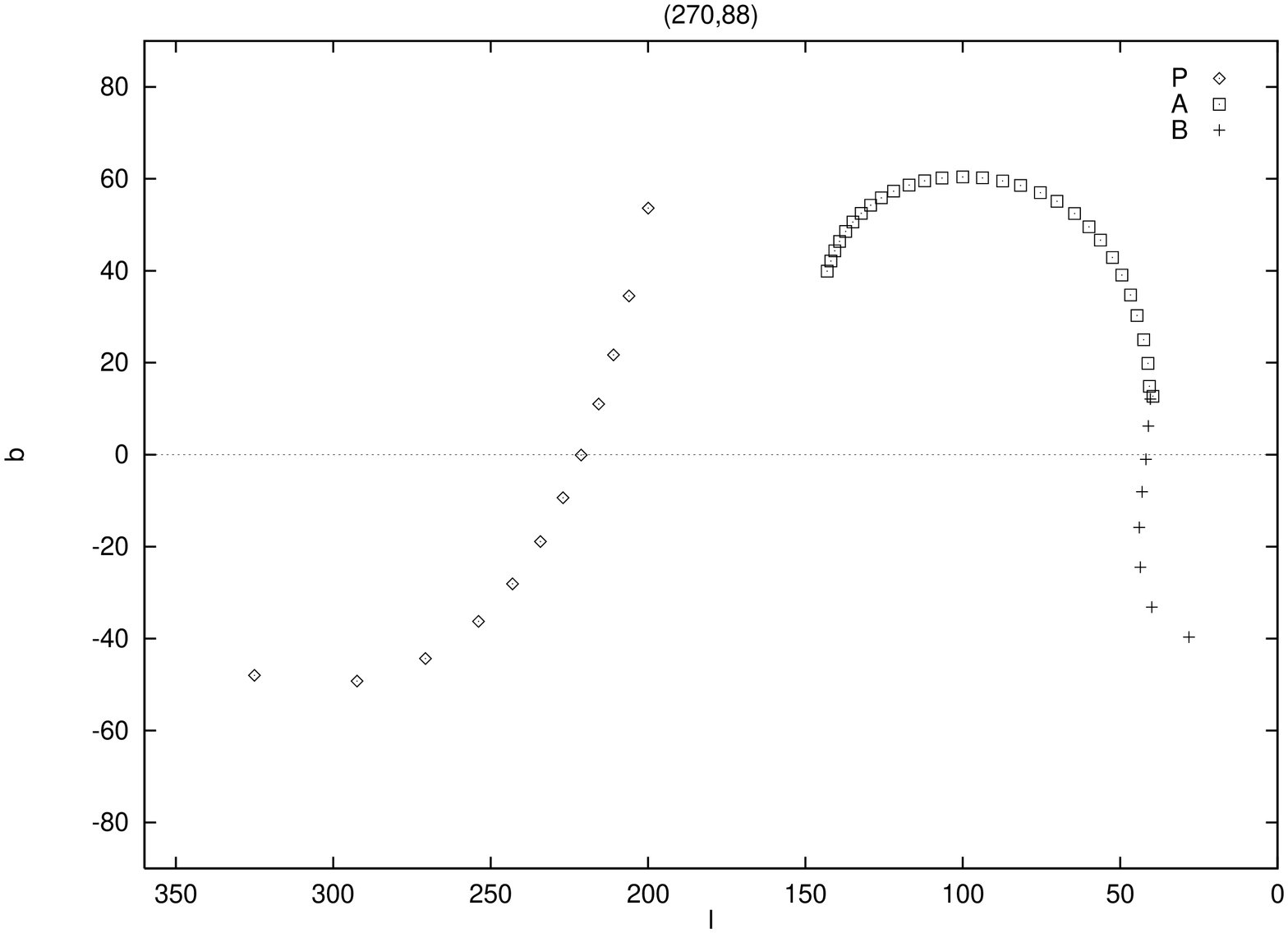,height=2.9in,angle=-90}}
\vspace{10pt}
\caption{Amplification vs. energy of the CR flux 
that enters the galactic wind from $(\ell,b)=(270^\circ,88^\circ)$
(left), and predicted arrival directions of 50 EHECR events with
$\overline E$ above 75~EeV assuming an injection flux proportional
to $E^{-2.7}$ (right).}
\label{fig4}
\end{figure}

The principal image is magnified by a factor of order 10 at 
$\overline E\approx 200$~EeV, is further amplified at intermediate
energies, and then its magnification starts to rapidly decrease
while $\overline E$ is still above 100~EeV. Its apparent position moves
south as the energy decreases. 
Secondary images appear at the energy at which the caustic
crosses the source position. 
One of the secondary images moves north and remains highly magnified, 
with an amplification factor above 100, while the other moves south
and is quickly demagnified. Notice that there are no events at 
southern galactic latitudes below a certain energy threshold. 

The fact  that magnification factors well above 100 are attained in a 
significant energy range, with $\overline E$ below 150~EeV, reduces the
energy requirements upon the source, that would need to be a factor of
more than 100 less powerful than if unlensed to provide the same 
observed flux in this energy range.

A  definite prediction of this model is a strong asymmetry
between events arriving from the northern and southern
galactic hemispheres. Although with the present EHE data, which
involves only the northern terrestrial hemisphere and hence mainly the
northern galactic one, it is not yet possible to test this
asymmetry,
the future operation of the Auger observatory, will allow to check 
the viability of this model. 
A latitude dependent upper cut-off value below
$2\times 10^{20}$~eV for CR protons arriving to the south and 
lower fluxes in the south than in the north above
$10^{20}$~eV are generic predictions of this scenario.

A galactic wind with a local value smaller than the 
7 $\mu$G adopted in \cite{ahn99}
could in any case have interesting observational consequences if EHECRs 
have a significant component which is not light. The flux of heavy nuclei 
could thus be strongly amplified by the galactic wind field, while the 
proton flux at the same energy would remain essentially unlensed. A transition 
to a heavy composition could thus result at extremely high energies.

\section*{Acknowledgments}

Work partially supported by ANPCyT, CONICET and Fundaci\'on Antorchas, 
Argentina.

\end{document}